# Nanowire quantum dots tuned to atomic resonances

*Lorenzo Leandro[1], Christine P. Gunnarsson[1], Rodion Reznik[2,3], Klaus D. Jöns[4], Igor Shtrom[2], Artem Khrebtov[3], Takeshi Kasama[1], Valery Zwiller[4,5], George Cirlin[2,3] and Nika Akopian*[1]*

[1]DTU Department of Photonics Engineering, Technical University of Denmark, 2800 Kgs. Lyngby, Denmark

[2]St.Petersburg Academic University, RAS, St. Petersburg 194021, Russia

[3]ITMO University, Kronverkskiy pr. 49, 197101 St. Petersburg, Russia

[4]Department of Applied Physics, KTH Royal Institute of Technology, SE-100 44, Stockholm, Sweden

[5]Kavli Institute of Nanoscience, TU Delft, 2628CJ Delft, Netherlands

*e-mail: nikaak@fotonik.dtu.dk

**Quantum dots tuned to atomic resonances represent an emerging field of hybrid quantum systems where the advantages of quantum dots and natural atoms can be combined. Embedding quantum dots in nanowires boosts these systems with a set of powerful possibilities, such as precise positioning of the emitters, excellent photon extraction efficiency and direct electrical contacting of quantum dots. Notably, nanowire structures can be grown on silicon substrates, allowing for a straightforward integration with silicon-based photonic devices.**

**In this work we show controlled growth of nanowire-quantum-dot structures on silicon, frequency tuned to atomic transitions. We grow GaAs quantum dots in AlGaAs nanowires with a nearly pure crystal structure and excellent optical properties. We precisely control the dimensions of quantum dots and their position inside nanowires, and demonstrate that the emission wavelength can be engineered over the range of at least 30 nm around 765 nm.**

**By applying an external magnetic field we are able to fine tune the emission frequency of our nanowire quantum dots to the $D_2$ transition of $^{87}$Rb. We use the Rb transitions to precisely measure the actual spectral linewidth of the photons emitted from a nanowire quantum dot to be 9.4 ± 0.7 μeV, under non-resonant excitation. Our work brings highly-desirable functionalities to quantum technologies, enabling, for instance, a realization of a quantum network, based on an arbitrary number of nanowire single-photon sources, all operating at the same frequency of an atomic transition.**

Exploiting the advantages of atomic and semiconducting systems opens new possibilities and research directions. For instance, quantum dots have been frequency tuned and locked to atomic resonances to provide a single photon source with stable emission frequency [1–3]. Such hybrid systems enable multiple emitters working at the exact same wavelength, which is one of the major challenges in quantum communication and quantum information processing. Recently, a different material system has been used to tune InGaAs quantum dots to Cs atoms, optically with a laser [4] and by electrically-induced strain [5,6]. Furthermore, electrically pumped quantum dots were developed to make the first quantum-LED tuned to atomic lines [7], paving the way to on-chip electrically controlled tunable single and entangled photon sources. In addition, coupling between a single quantum dot and a single ion has been shown [8], allowing for direct transfer of quantum information from a single quantum dot to a single atom.

Nevertheless, all previous works have used quantum dots in materials grown in bulk, while embedding the quantum dots in nanowires provide additional important functionalities to the system, making them one of the most promising systems for numerous applications in quantum nanophotonics [9–13]. Over the last few years several groups have successfully demonstrated nanowire quantum dots in various material systems [14,15]. Recently, the exciton energy of single nanowire quantum dots on piezo crystals was tuned into resonance by applying strain [16], and Bell´s inequality violation [17] and entanglement [18,19] have been also demonstrated. The advantages of the nanowire structure are, among others, high extraction efficiency [10,11,20,21], the possibility of maximizing the coupling to optical fibers [10,22,23], the positioning of nanowires in highly uniform arrays with higher uniformity than what is achievable with self-assembled quantum dots [24–28], and the deterministic and scalable integration with photonic nanostructures that can lead to on-chip photonic circuits [16,29,30].

GaAs quantum dots in AlGaAs structures are our preferred material system - their emission energy can be engineered close to the $D_2$ transitions of $^{87}$Rb [1], they are the most pure single photon sources reported to date [31], and they show the highest degree of polarization entanglement of photons [32] generated from semiconductor quantum dots. Furthermore, GaAs quantum dots grown in AlGaAs nanowires are expected to have much sharper interfaces between the two materials, since aluminum crystallizes at least 100 times faster than gallium [33,34], quickly leaving the droplet without aluminum once the source is closed. This allows for a higher control of the quantum dot shape, compared to other material systems. The first works have already shown the growth of GaAs quantum dots in AlGaAs nanowires [35,36]. However, the optical properties were not yet optimal to address quantum applications, for instance, the emission linewidth spans from 95 µeV [35] to 2 meV [36]. As of today, realization of a controlled growth of high optical quality nanowire quantum dots working at atomic frequencies requires development of the material system. Here, we demonstrate precisely engineered GaAs quantum dots in AlGaAs nanowires with nearly-pure crystal structure, high optical quality, and frequency-tuned to the optical transition of Rb atoms.

We grow our GaAs quantum dots in AlGaAs nanowires on a Si wafer using the Vapor Liquid Solid technique in a Molecular Beam Epitaxy reactor, using a gold particle as catalyst. The growth takes place just below the catalyst particle, where the atoms from the vapor crystallize, forming the nanowire. To grow a quantum dot we stop the Al source for a few seconds (e.g. 5 s, 7 s or 15 s). This brief growth of only GaAs creates a quantum dot confined between AlGaAs segments of the nanowire. We set the growth parameters to allow for a simultaneous axial and radial growth, resulting in tapered nanowires (for details see methods section). The radial growth results in an essential shell that protects the quantum dot from the environment outside the nanowire. We grow several sets of nanowires, with heights in the range of 3-6 μm and with diameters (core plus shell) in the range of 150-200 nm at the position of a quantum dot, depending on the growth conditions. The quantum dots have diameters defined by the size of the gold catalysts, which is about 20 nm.

In Figure 1a we present a schematics showing the quantum dot and the core-shell structure of the nanowire. Figure 1b and 1c depict 2D energy-dispersive X-ray spectroscopy (EDS) elemental maps for Ga and Al, respectively. As expected, we observe an increase in the Ga content with a corresponding decrease of Al content at the quantum dot position. To clearly show the quantum dot, in Figure 1d we present 1D EDS scans along the nanowire core. We note that the expected sharp interfaces are not observed here due to the small concentration difference between GaAs and AlGaAs and a beam broadening inside the material which reduces the sensitivity and the spatial resolution of the measurement.

In Figure 2a we show the photoluminescence (PL) spectra at 4.2 K from three quantum dots, grown for different times: 5 s, 7 s and 15 s. As we expect, increase in the growth time increases the height of the quantum dot, which decreases the emission energy (increases the emission wavelength) of the quantum dot. As we show in Figure 2b, the PL spectra of our nanowire quantum dots has narrow linewidth, limited by the resolution of our spectrometer (30 μeV). In Figure 2c we plot an autocorrelation measurements of the exciton emission of a quantum dot, showing antibunching with a fitted $g^{(2)}(\tau=0) = 0.19$ (see SI section 4). Our results show that we can engineer the wavelength of the emission from our nanowire quantum dots over a range of 30 nm around 765 nm, by selecting the growth time from 5 s (emission at 750 nm) to 15 s (emission at 780 nm). Our nanowire quantum dots are spectrally narrow and bright sources of single photons, with rates of 2.2 Mcounts/s under continuous wave excitation (see SI section 3), even without any intentional cavity integration. The count rates can be further increased, for instance, by implementing mirrors at the bottom of the nanowires [37,38] and by optimizing the waveguide design of the nanowire [11,39]. The narrow and intense lines also suggest that our nanowires are defect free and of a very high crystal purity, in some cases down to only 10 stacking faults per micrometer (see SI section 1). We note, however, that not all of our nanowires are free from defects, and not all quantum dots have excellent optical properties.

Even if one can precisely control the dimensions of quantum dots, no growth method allows obtaining systematically identical heterostructure quantum dots. In virtually all cases, the emission frequency is slightly different from one quantum dot to the other, not allowing for exchange of photons between distant quantum dots. Therefore, a reliable post-growth technique is necessary to fine-tune each quantum dot to a frequency reference provided, for instance, by an atomic transition line. To demonstrate such tuning with our nanowire quantum dots, we interface them with a vapor of Rb atoms, similarly to our earlier work on quantum dots grown in a multi-step self-assembly process in bulk [1]. In the absence of strong electric or magnetic fields, these atomic transition energies are the same, independently on time, location and number of emitters.

We use the $D_2$ optical transition of $^{87}$Rb atoms as established frequency references to which we fine-tune the emission from our nanowire quantum dots, using an external magnetic field. This is done by means of the Zeeman effect and diamagnetic shift. During the experiment, we excite the nanowire quantum dot with a laser and then direct its emission through a cell with a vapor of $^{87}$Rb (see Figure S11 for schematics). Varying the applied magnetic field, we change the frequency of the emitted photons. The photons propagate through the Rb cell and as their frequency matches the atomic transitions, they get absorbed or scattered. The transmission of photons through the Rb cell is then measured.

In Figure 3, we show the results of our experiment. We scan the magnetic field from 0.7 T to 0 T and then back to 0.7 T for reproducibility. We then trace the short-wavelength branch of the Zeeman-split lines (highlighted by dashed lines in Figure 3a) and analyze the transmitted intensity, shown on Figure 3b. We clearly resolve two transmission dips due to the hyperfine structure of the $D_2$ lines of $^{87}$Rb. Since the experiment is repeated twice (from 0.7 T to 0 T and from 0 T to 0.7 T), Figure 3b shows four dips. We then fit the data to our model [1] (see SI section 6) and take advantage of our hybrid system to perform high-resolution spectroscopy of our nanowire quantum dot, similarly to laser spectroscopy [3]. The fit, which has only one fitting parameter – the spectral linewidth of the emission – reveals a narrow exciton line of 9.4 μeV. To the best of our knowledge, this is the narrowest linewidth reported for GaAs nanowire quantum dots [15,40,41], and can be further improved by exciting the quantum dots resonantly.

In summary, we developed a technique for controlled growth of GaAs quantum dots in AlGaAs nanowires on Si substrates, where we precisely engineer the size and position of the dots inside the nanowire. Our nanowires are nearly defect free, and we can design the quantum dot to emit single photons anywhere in the range of 30 nm around 765 nm. These nanowire quantum dots are bright emitters with very narrow spectral linewidth, down to 9.4 µeV. We demonstrated the tuning of their emission to the $D_2$ lines of $^{87}$Rb, enabling operation of our nanowire quantum dots at a stable and precise reference frequency. The hybrid system we present in this work combines the reliability and stability of natural atoms with a variety of functional advantages of nanowire quantum dots. Nanowire quantum dots tuned to atomic resonances could be one of the key elements in quantum information networks, on-chip photonic circuits, and in generation of multi-dimensional photonic cluster states [42].

**Methods**

We grow our nanowires on a Si(111) wafer. The wafer is first outgassed at 850 °C and then the Au is deposited with total thickness of 0.1 nm at 550 °C. After deposition, the Au layer will form droplets of about 20 nm in diameter, as this formation minimizes the potential energy. In order to improve the homogeneity of the Au droplet size, the substrate is kept for 1 min at 550 °C - this time is long enough to ensure that the droplets become similar in size. Afterwards the substrate is cooled to room temperature and transferred to the growth chamber with no vacuum brake. During the nanowire growth the substrate temperature is set at 510 °C, the V/III flux ratio is 3, and the total AlGaAs growth rate is fixed at 0.3 nm/s, for all samples in this work. The AlGaAs part of the nanowire is grown for 20 min or 25 min, depending on the desired height of the final nanowire, and then the Al source is closed for either 5, 7 or 15 sec in order to form the GaAs quantum dots. Afterwards, the supply of Al is opened again for 5 min to continue the nanowire growth. The growth method results in tapered core-shell nanowires (see Fig. 1a). The Au droplet diameter will determine the nanowire core diameter, and hence the quantum dot diameter. As the distribution of the size of the Au droplets is ensured to be homogeneous, the nanowire core and the quantum dot for all samples will have a diameter close to 20 nm. The diameter of the total (core+shell) nanowire is in the range 150–200 nm at the position of the quantum dot, depending on the total AlGaAs growth time. The height of the quantum dot inside the nanowire is given by the time in which the Al shutter is closed (either 5, 7 or 15 sec in these samples). The Al shutter closing time is 0.3 s and thus allows accurate control the quantum dot height. The density of the nanowires was kept low during growth (see Figure S1), in order to be able to distinguish individual nanowire quantum dots.

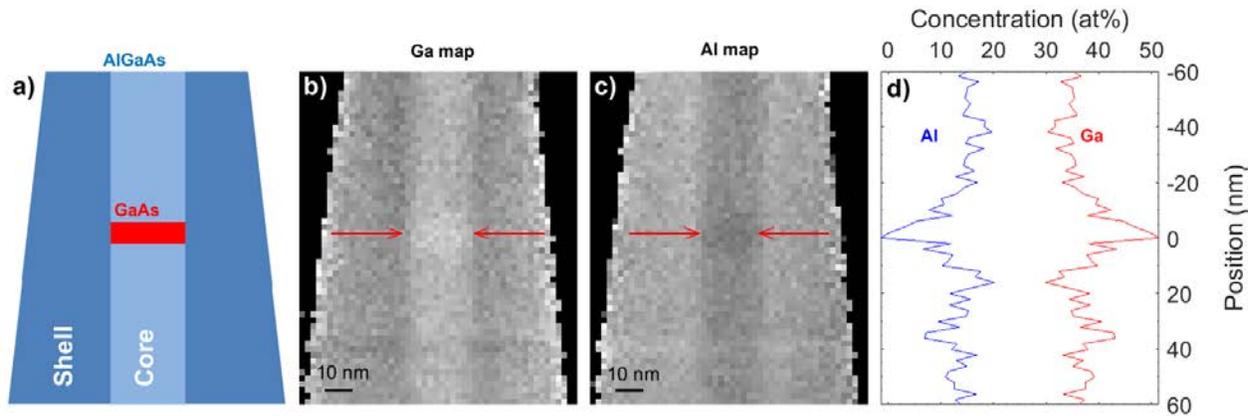

**Figure 1. GaAs quantum dots in AlGaAs nanowires. a)** Schematics of a GaAs quantum dot (red) in a AlGaAs nanowire (blue), composed of a core surrounded by a shell. **b)** and **c)** Concentration maps of Ga and Al in a nanowire measured using STEM-EDS, respectively. The quantum dot is seen as an increase in Ga concentration and a corresponding decrease in Al concentration. **d)** Concentrations of Ga (red) and Al (blue) in the nanowire core along its length, corrected for imperfect measurement technique and low material contrast (see section SI 1). All measurement in this figure are done on a sample specifically grown to have the quantum dot closer to the top of the nanowire to reduce the effect of the shell on the measurements (see section SI 1).

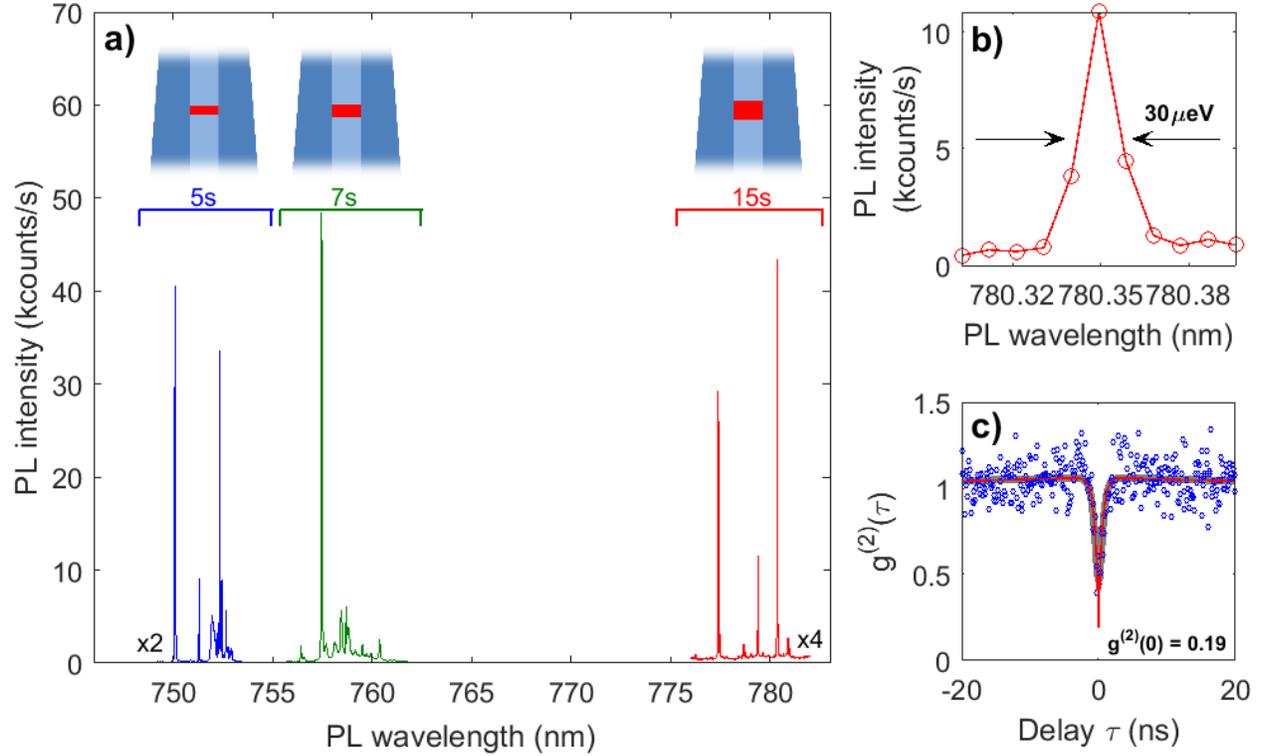

**Figure 2. Optical characterization of nanowire quantum dots. a)** Photoluminescence (PL) spectra at 4.2 K of three samples with different growth times of the quantum dot − 5, 7 and 15 s (for more details see SI section 3). A schematic of the quantum dot in nanowire structure is shown above each corresponding spectra. **b)** A spectral zoom on the exciton line at ~780 nm ($D_2$ line of $^{87}$Rb). **c)** Autocorrelation measurements of an exciton (750 nm) in a nanowire quantum dot (5 s). The blue circles are the measured data and the grey line is the fit, which takes into account the finite time resolution of our detectors. The red line shows how the measurement would appear for an ideal system response (for more details see SI section 4).

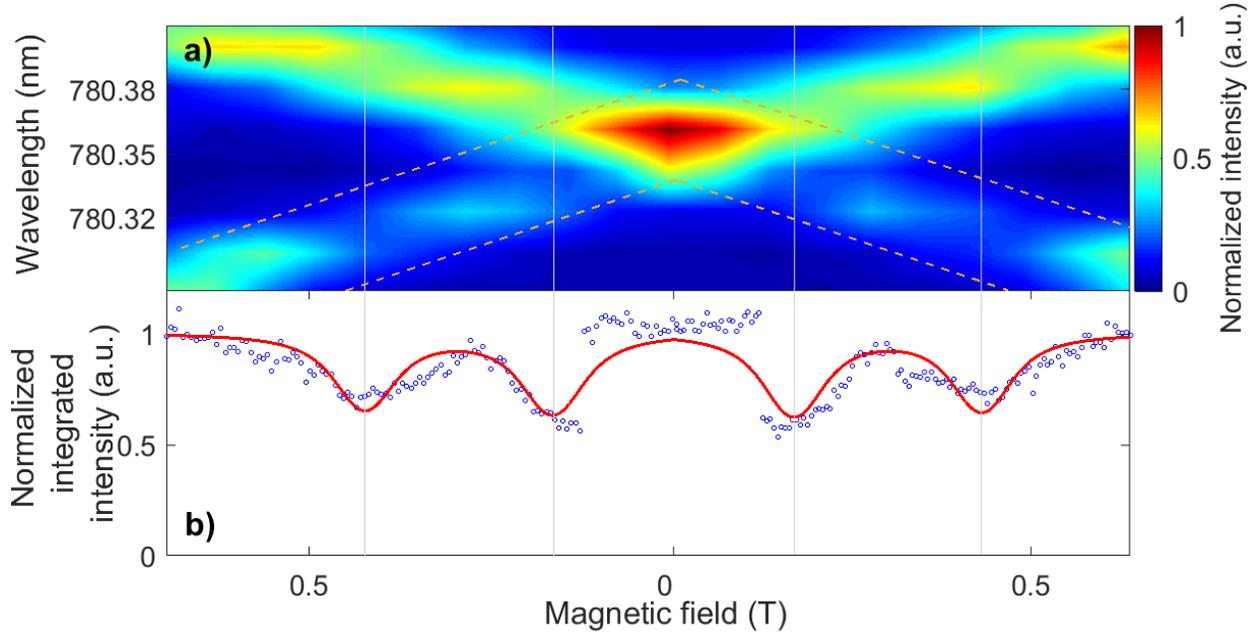

**Figure 3. Transmission through Rb vapour.** **a)** PL spectra of an exciton in a nanowire quantum dot scanned through the atomic transitions. The experiment is done by swiping the magnetic field from 0.7 T to 0 T and then back to 0.7 T. Colors represent the normalized PL intensity. The data is then interpolated for clarity (see raw data in Fig. S12). **b)** Transmission of the exciton line through the Rb vapor, obtained by tracing the short wavelength branch of the Zeeman-split exciton in the raw data, marked by dashed lines in a). Blue points refer to the experimental data, while the red line represents the fit to our model. The fit reveals a very narrow emission linewidth of 9.4 ± 0.7 µeV.

**Supporting information**

Includes electron microscopy images of the samples, further optical characterization measurements, fitting procedures and details on data interpolation used in the main text.

**Acknowledgements**

We gratefully acknowledge the support of Villum Fonden (Project no. VKR023444).

The nanowire samples were grown under the support of the Russian Science Foundation (Project no 14-12-00393).

We gratefully acknowledge CryoVac GmbH (www.cryovac.de) for developing a cryogenic system used in some of the experiments reported here.

# Supporting Information

# Nanowire quantum dots tuned to atomic resonances

*Lorenzo Leandro[1], Christine P. Gunnarsson[1], Rodion Reznik[2,3], Klaus D. Jöns[4], Igor Shtrom[2], Artem Khrebtov[3], Takeshi Kasama[1], Valery Zwiller[4,5], George Cirlin[2,3] and Nika Akopian*[1]*

[1]DTU Department of Photonics Engineering, Technical University of Denmark, 2800 Kgs. Lyngby, Denmark

[2]St.Petersburg Academic University, RAS, St. Petersburg 194021, Russia

[3]ITMO University, Kronverkskiy pr. 49, 197101 St. Petersburg, Russia

[4]Department of Applied Physics, KTH Royal Institute of Technology, SE-100 44, Stockholm, Sweden

[5]Kavli Institute of Nanoscience, TU Delft, 2628CJ Delft, Netherlands

*e-mail: nikaak@fotonik.dtu.dk

# Contents

# 1. Scanning and Transmission Electron Microscopy

In Fig. S1, we show Scanning Electron Microscopy images, top and side view, to show the density of nanowires on the substrate and the final nanowire shape. In Fig. S2 we show our high-angle annular dark-field (HAADF) measurement which depicts the core and the shell of the nanowire, distinguished because of their contrast difference. The contrast difference between the core and the shell is due to their different Al content. The quantum dot is visible as brighter spot at the center of the images, especially noticeable in Fig. S2(right). This latter figure, together with Fig. 1 in the main text, contains measurement performed on a sample specifically grown to have the quantum dot higher up in the nanowire, in the tapered part close to the end. This is done to be able to see the quantum dot through the shell. Also, this sample was grown with higher Al molar fraction (0.6), to further increase the contrast between a quantum dot (GaAs) and a nanowire (AlGaAs). The quantum dot growth time was 15 seconds.

The Al content is higher in the shell than in the core because of the suppressed diffusivity of the Al atoms on the sidewalls of the nanowire 1. Further Transmission Electron Microscopy measurements show a very low density of structural defects in the entire nanowire structure. An example of such measurement is shown in Fig. S3. Here we demonstrate a density of stacking faults of about 10 per µm.

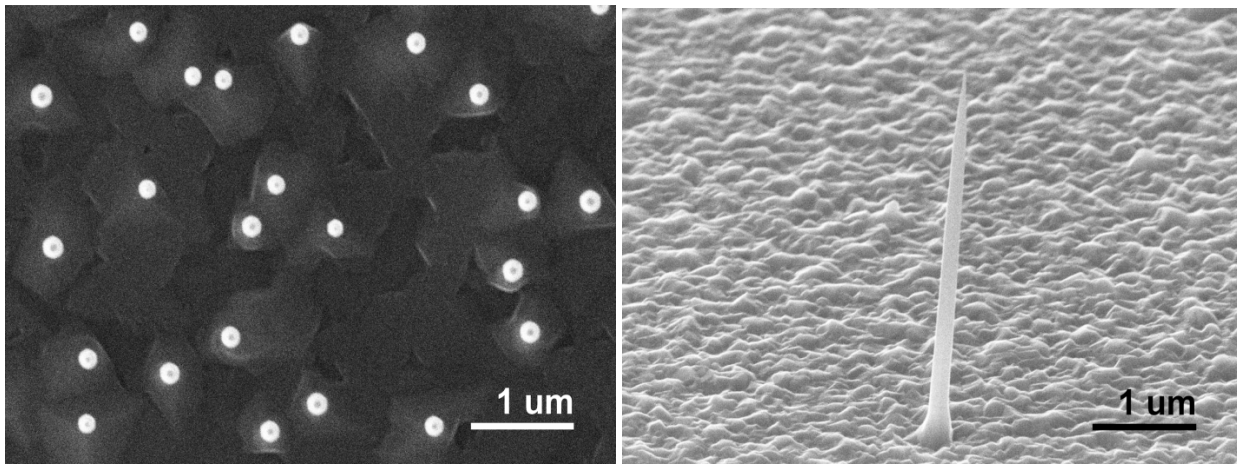

**Figure S1 | Scanning Electron Microscopy.** Scanning Electron Microscopy images showing nanowire quantum dots from a top view (left) and one individual wire from side view (right).

We have shown that our growth method allows us to control the height of the quantum dot, which is done by changing the growth time of the GaAs section in the nanowire. The height is the smallest dimension in the quantum dot, hence it will define the emission wavelength. The lateral dimension is given by the diameter of the Au droplet, which have a small distribution around 20 nm. A small variation in the diameter of the quantum dot will then give a narrow distribution of emission wavelengths, but this effect is much weaker than the effect of the quantum dot height on the emission wavelength. Thus our growth procedure enables control of the nanowire quantum dot size, and results in high optical quality samples, not previously reported in the literature for GaAs quantum dots in nanowires[2–4].

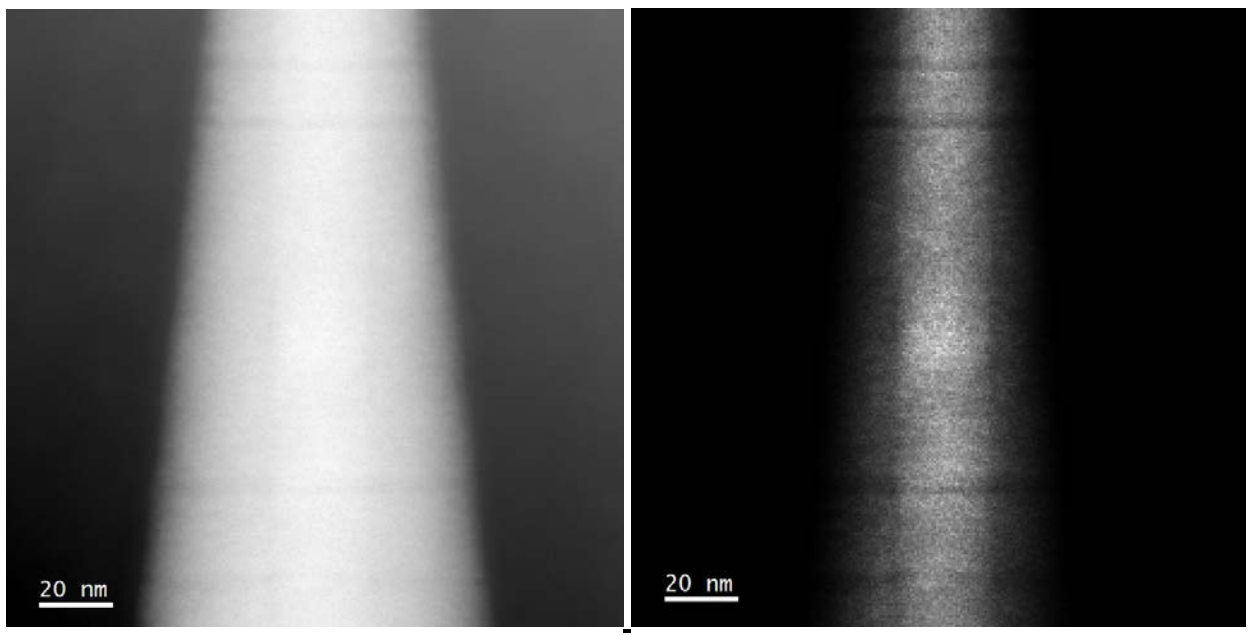

**Figure S2 | Scanning Transmission Electron Microscopy.** Scanning Transmission Electron Microscopy images taken with an high angle annular dark field detector, from the same measurement but with different contrast to show a nanowire quantum dot with low density of stacking faults. Different brightness enables to distinguish the core and the shell of the nanowire, but also the quantum dot which is visible as brighter area in the center of the image. Stacking faults appear as dark horizontal lines. This sample is specifically grown to have the quantum dot closer to the top of the nanowire to reduce the effect of the shell in the measurement.

S4

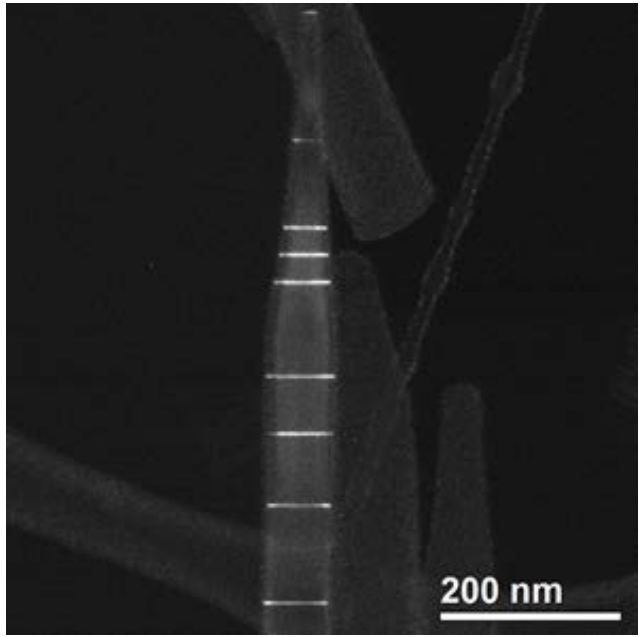

**Figure S3 | Dark Field Transmission Electron Microscopy.** Dark Field Transmission Electron Microscopy image showing the density of stacking faults in a Al0.4Ga0.6As nanowire.

## 2. Energy-dispersive X-ray Spectroscopy (EDS) line scan corrections

We did EDS measurements on the nanowires to determine the amount of Al and Ga in them. From the EDS line scan shown in Fig SI 2.1a, we corrected the concentration of Ga and Al to sum up to 50%, assuming that Ga and Al occupy the same atomic sites. Furthermore, to reduce the electron channeling effect, the nanowire samples were slightly tilted from a strong diffraction condition 5. Thus we write:

$$C_i^{corr} = C_i^{meas} * \frac{50}{C_{Al}^{meas} + C_{Ga}^{meas}} \quad (S1)$$

Where $C_i^{meas}$ ($i = Al$ or $Ga$) are the measured atomic concentrations and $C_i^{corr}$ are the corrected atomic concentrations, in at%. Another thing that affects the measured concentration in the core is the atomic concentration in the shell, which is also measured. The effect of the shell can be removed by considering the concentration change given by the shell as a linear effect:

$$C_i^{corr} = C_i^{core,meas} * \frac{d_{core}}{d_{total}} + C_i^{shell,meas} * \frac{2*d_{shell}}{d_{total}} \quad (S2)$$

where $d_{total} = d_{core} + 2 * d_{shell}$ (see Fig. S5). Hence we obtain the concentration of the core by rewriting:

$$C_i^{core} = \frac{d_{total}}{d_{core}} * \left[ C_i^{corr} - C_i^{shell} * \frac{2*d_{shell}}{d_{total}} \right] \quad (S3)$$

The concentration in the shell $C_i^{shell}$ is obtained by measuring only along the shell, averaging the measurements and accounting for the channeling effect using Eq. S1. The data, after applying the two formulas, is shown in Fig. S4b, where the Ga concentration mirrors the Al, because of equation S1. All measurements and schematics in this section refer to the sample specifically grown for Transmission Electron Microscopy discussed in section SI 1.

The fact that the shells influence on the measurement of the core can be considered as a linear effect is understood by considering that the electron beam is small in size compared to the change in the shape of the nanowire, as depicted in Fig. S5. This means that, in first approximation, the volume probed by the electron beam resembles a perfect cylinder containing the shell, the core and the shell again. Since the diameter of the cylinder is assumed the same for the core and the shell, the ratio of the volume probed by the electron beam, $\frac{V_{core}}{V_{shell}}$ simplifies to $\frac{d_{core}}{d_{shell}}$.

We expect higher interface sharpness for AlGaAs-GaAs material system in comparison with other materials, because the aluminum crystallizes much faster than gallium, leaving quickly the droplet without aluminum once the source is closed[1]. This is a significant advantage when it comes to designing complex structures, such as multiple quantum dots in a nanowire. We note, however, that the resolution of the EDS measurement for GaAs/AlGaAs is not high enough to clearly show the quantum dots sharpness, because of the small concentration difference between Ga and Al.

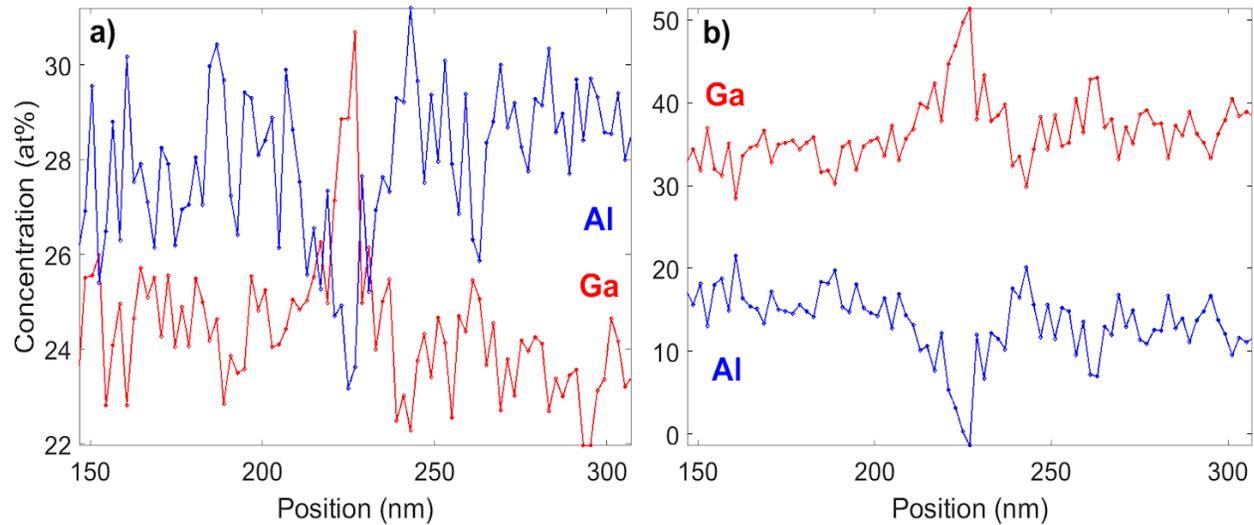

**Figure S4 | Data analysis of EDS scans. a)** The raw data of the atomic concentration of Al along the nanowire in the vicinity of the quantum dot. **b)** The atomic concentration data after the corrections described in section SI 2.

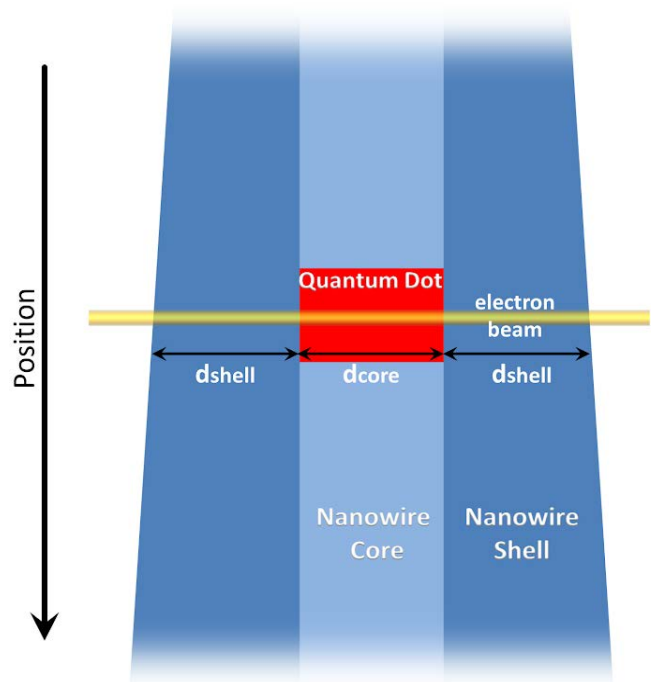

**Figure S5 | Schematics of a quantum dot in nanowire.** The electron beams enters the nanowire probing both the shell and the quantum dot in the core. The size of the beam is assumed uniform across the whole nanowire.

## 3. Optical Characterization

All optical experiments were performed in a He bath cryostat, at a temperature of 4.2 K, containing a 9 T superconducting magnet. The quantum dots were excited above bandgap with a continuous-wave laser at 532 nm. The energy levels of the confined system, which determine the emission energy, are primarily given by the height of the quantum dot. Using a microscope objective (NA=0.85), the laser was focused to a spot size of about 1 µm. The quantum dot excitation light and photoluminescence were both collected through the same objective. The photoluminescence was spectrally filtered by a spectrometer to a charged-coupled device (CCD) or an avalanche photodetector (APD). The measured total setup efficiency is approximately 1.7 %. We thus calculate the brightness of our sources under CW excitation, for instance, from the emission line at 750 nm (figure 2a in the main text):

$$\frac{37\frac{kcts}{s}}{1.7} = 2.2\ Mcts/s \tag{S4}$$

**Scanning photoluminescence**

To clearly distinguish the emission lines coming from the quantum dot and the ones coming from the nanowire, we performed photoluminescence measurements on specially grown nanowire samples with and without quantum dots. The Al molar fraction in the nanowires was increased to 0.6 and the quantum dot growth time was increased to 22 s, in order to enlarge the spectral separation between the emission from the nanowire and the quantum dot.

To demonstrate a statistical distribution of the emission from these samples we performed the measurements by continuously integrating the photoluminescence for 30 s, while the sample is scanned. This allows to obtain a measurement integrated over a large number of nanowires, similarly to what can be achieved by macro-photoluminescence. We refer to this technique as scanning photoluminescence.

In Fig. S6 we show the resulting spectra for the reference sample with no quantum dots (Fig. S6a), and for a sample with quantum dots (Fig. S6b). The comparison between the two measurements clearly identifies the emission from the quantum dots, showing a statistical distribution of the emission of different quantum dots of ~10 nm full width half maximum. From Fig. S6, it is also clear that the nanowire emission is composed by two peaks. This is attributed to the different Al concentration in the core in respect to the shell, and is currently under our further investigation.

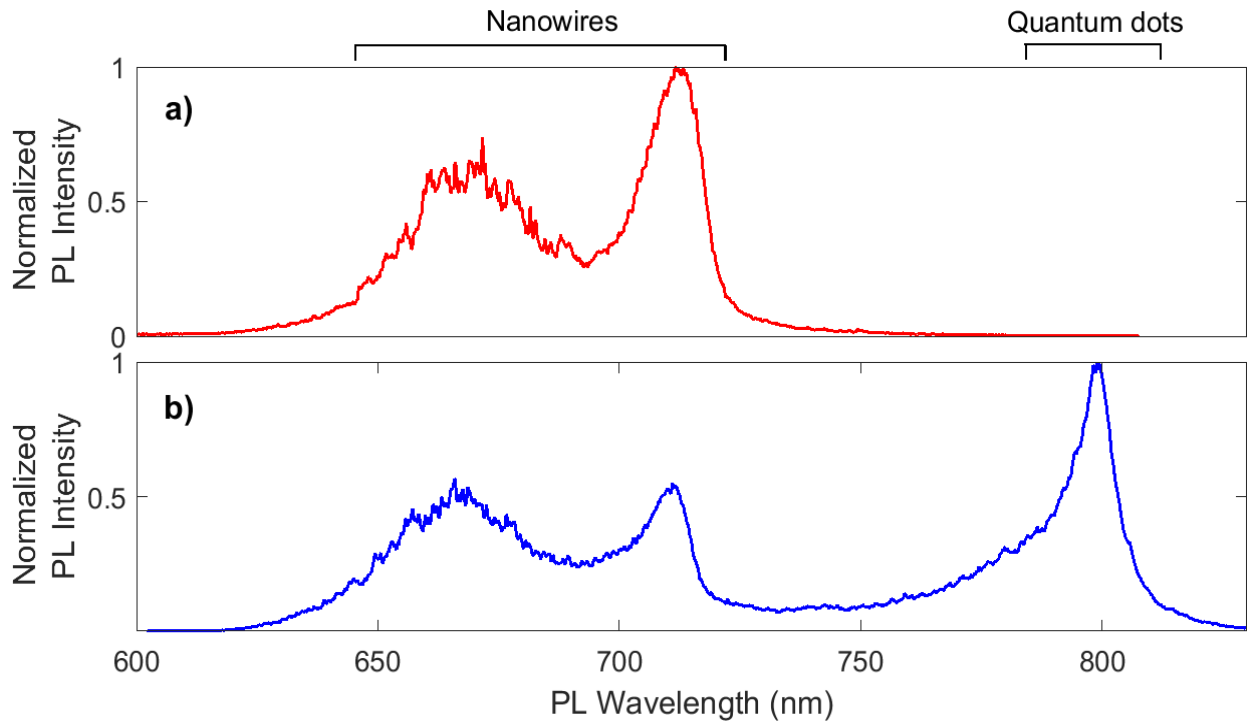

**Figure S6 | Scanning photoluminescence.** Scanning photoluminescence spectra of high-density samples made with 0.6 aluminum molar fraction measured scanning over a large number of nanowires. **a)** Spectrum from a reference sample without quantum dots where only the emission from the nanowire is visible. **b)** Spectrum from a sample with a quantum dot grown for 22 s. In the latter, the two nanowire peaks are still visible while the quantum dot emission appears at higher wavelength.

**Excitation power dependent photoluminescence**

In Fig. S7 we report an example of power dependent photoluminescence for the 5 s quantum dot under continuous-wave excitation. The spectrum of the same quantum dot, at 400 nW excitation power, is also shown in Fig. 2a in the main text. In Fig. S7a we plot the spectrum evolution under different excitation powers, while in Fig. S7b we show the tracing of several of the lines appearing in the spectrum. Their power dependence shows a linear slope followed by saturation and subsequent intensity reduction, which is what to be expected from excitonic complexes under continuous-wave excitation. The three brightest lines (A, B and D) show an equal slope close to unity, indicating that they are excitons (e.g. neutral or charged). Other lines (C, E and F) show superlinear power behavior, which point at biexcitonic origin of the emission, although none of these lines has shown the expected slope of 2. Nevertheless, given the rich emission spectrum, this behavior can be attributed to competing charging mechanisms under a non-resonant excitation.

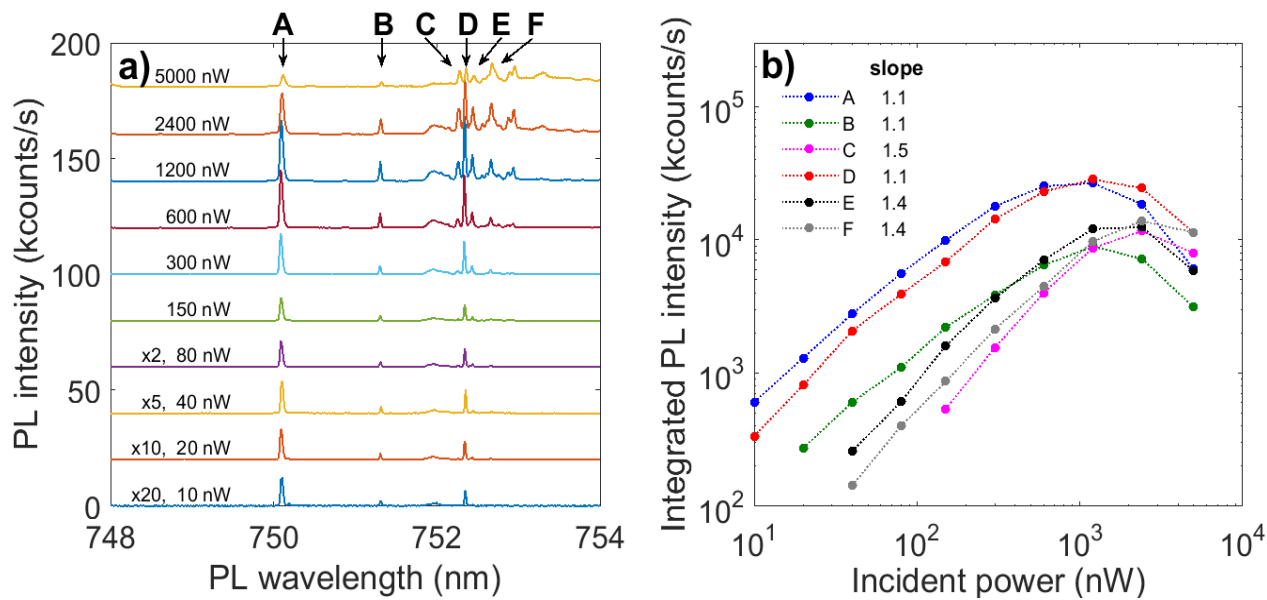

**Figure S7 | Excitation power-dependent photoluminescence. a)** Photoluminescence spectra under a non-resonant continuous-wave excitation at different powers, at 4.2 K, of the sample with quantum dots grown for 5 s, corresponding to the one shown in Fig. 2a in the main text. **b)** Traced integrated peaks (5 pixels around maxima) of lines indicated in a) showing a linear slope in log scale and saturation at high excitation power.

**Intensity correlations measurements**

In Fig. S8, we show intensity correlation measurements of line A with itself and the two other bright lines in the spectrum – B and D. These measurements show that all these lines are mutually exclusive, demonstrating that they originate from the same quantum dot.

The cross-correlation measurement in Fig. S8b shows slight bunching that can be expected from correlations between excitons with different charge configurations (such as charged and neutral excitons, see for example Fig. S3 of reference[6]). The slightly asymmetric antibunching in Fig. S8c suggests that lines A and D also originate from states with different charge configurations.

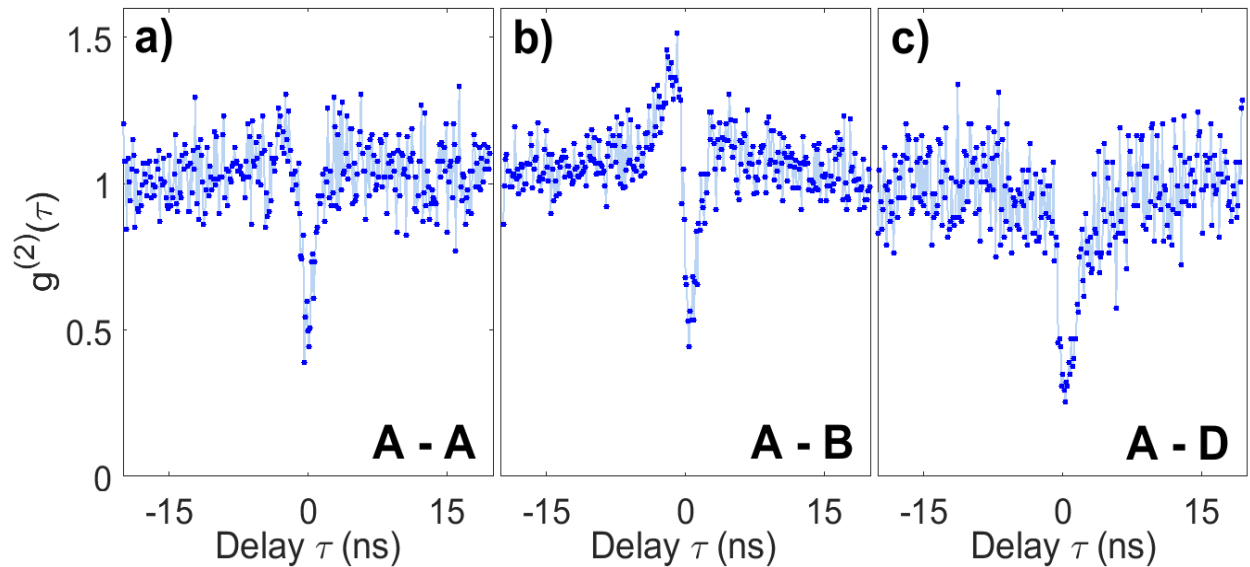

**Figure S8 | Intensity correlation measurements. a)** Auto-correlation measurement of line A (lines A, B and D as in Fig. S7), the same measurement as in Fig. 2a in the main text, shown here for reference. **b-c)** Cross-correlation measurements of the line A with the other two intense lines in the spectrum, line B and line D.

## 4. Fitting correlation measurements

We use a Hanbury Brown Twiss interferometer to measure temporal intensity autocorrelations of the light emitted from the quantum dot. The second order autocorrelation function is given by[7]:

$$g^{(2)}(t) = 1 + \alpha^2 * [\beta e^{-\frac{(t-x_o)}{\tau_b}} - (1+\beta) * e^{-(t-x_o)/\tau_a}] \quad \quad (S6)$$

where $t$ is the time, $x_0$ is the time offset from zero, $\alpha$ is the background light, $\tau_a$ is the lifetime of the quantum dot exciton, $\beta$ is the bunching coefficient, and $\tau_b$ is the effective lifetime of the bunching effect. The measured coincidence data is fitted to the second order correlation function $g^{(2)}$ convoluted with a Gaussian function, to take into account the finite response time of the detectors, which is ~600 ps. In Fig. S9 the data is shown by the blue dots and the grey curve shows the fit. From this fit we obtain $g^{(2)}(0) = 0.44$. The red curve shows the real (ideal response function) second order autocorrelation function (Eq. S6), calculated using the values found from the fit, and resulting in $g^{(2)}(0) = 0.19$. The non-zero value of the $g^{(2)}(0)$ is mainly due to the remaining background signal entering the single-photon detectors and their dark counts.

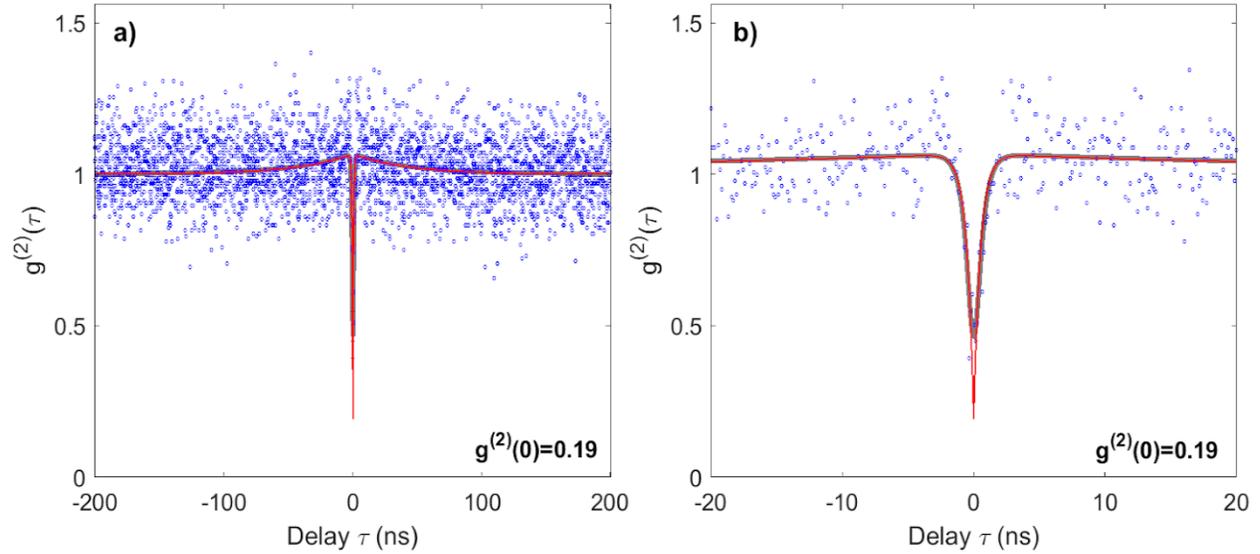

**Figure S9 | Auto-correlation measurement. a)** Autocorrelation measurements of an exciton (750 nm line) in a nanowire quantum dot (5 s). The blue circles are the measured data and the grey line is the fit, which takes into account the finite time resolution of our detectors. The red line shows how the measurement would appear for an ideal system response. **b)** same as a) but zoomed in to show the result of the measurement and the fit around $\tau$=0. Also shown in Fig. 2c of the main text.

## 5. Magneto-luminescence

In Fig. S10, we show the magneto-luminescence data for a quantum dot grown for 15 s, together with the fitting of the three main peaks. By this fitting to a quadratic polynomial, we calculated the g (linear) and γ (quadratic) coefficients, displayed in Table S1, using the formula:

$$E = E_0 \pm g m_j \mu_B B + \gamma B^2 \tag{S7}$$

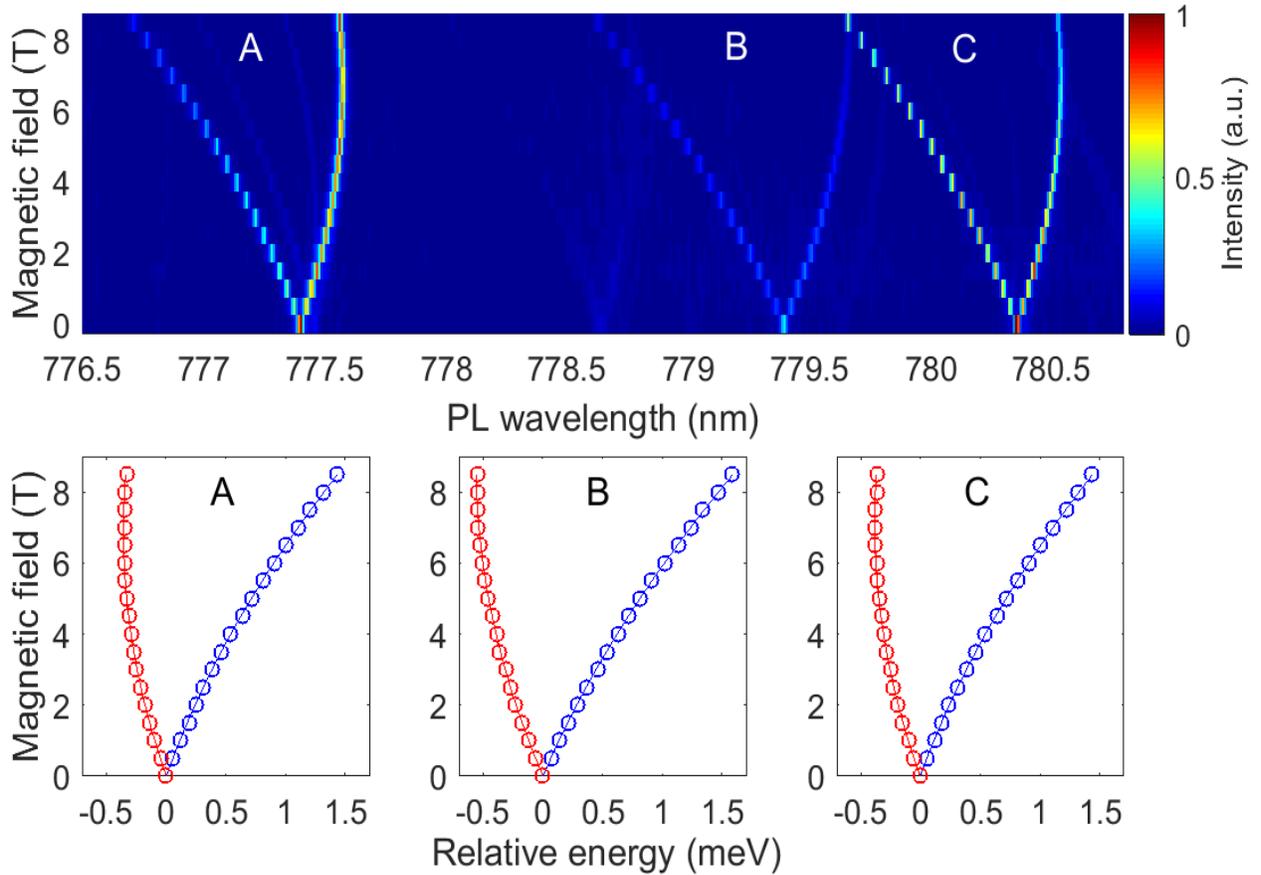

**Figure S10 | Magneto-luminescence.** Photoluminescence spectrum under application of an external magnetic field of a nanowire quantum dot grown for 15s. The emission lines split and shift due to the Zeeman effect (linear) and diamagnetic shift (quadratic). The spectra for one of the quantum dots is shown (top) with the fitting of the three main peaks (bottom, labelled A, B and C), where the circles represent the traced peak maximum and the lines show the best quadratic fit to the maxima (refer to equation S7). In the bottom graphs, the zero energy refers to the position of the peak at zero field.

| Peaks | \|g\| | γ [μeV/T$^2$] |
|---|---|---|
| A | 1.80 | 7.56 |
| B | 2.18 | 7.13 |
| C | 1.84 | 7.44 |

**Table S1 | Zeeman Coefficients.** The g and γ coefficients obtained by fitting the lines as in Fig. S10.

# 6. Tuning to rubidium, tracing of spectral line and interpolation of data

In Fig. S11 we show the relevant energy levels of our hybrid system of nanowire quantum dots interfaced with natural atoms, and the diagram of the experimental setup we used to tune nanowire quantum dots to atomic resonances. During the experiment, the transmission of the quantum dot emission through the atomic cloud is measured. These photoluminescence spectra are shown in Fig. 3a of the main text are interpolated (2D interpolation, see Fig. S12) to display better the absorption dips appearing when the quantum dot emission crosses the absorption lines of the Rb atoms. We used the raw data to trace the emission line, and to measure and fit its transmission through the cell. From the raw data we traced the short wavelength branch of the quantum dot emission, marked with orange dashed lines. This is done by following a quadratic behavior (assumed linear in such small range of applied external magnetic field) of the emission energy versus magnetic field, and integrating the intensity over 3 adjacent pixels. Once we obtained the transmission data points, we fitted them (red line in Fig. 3b of the main text) with a model, previously developed[8]. This fit is done with only one free parameter - the linewidth of the emission line. The other parameters are fixed to the actual experimental values.

We note that:

A. The apparent fluctuation in the intensity of the short wavelength branch (Fig. S12 a and b) result from the pixelated measurement of the emission line and not actual intensity changes;

B. The transmission data points (blue circles in Fig. 3b of the main text) show a discontinuity close to the zero field (~0.12 T). This results from one of the rubidium absorption lines being very close to the zero-field emission of the quantum dot in this particular case. Thus, near the zero we see both Zeeman lines together and as soon as we apply a small magnetic field one lines moves away while the other gets quickly absorbed. This effect is not general and it is not visible when the initial zero-field line is not so close to the absorption lines (see ref[8] as example);

C. The full width half maximum of the dips in the red fit line in Fig. 3b of the main text is not a direct measure of the linewidth of the emitter. The linewidth is retrieved by the fit, as only fit parameter, while considering the absorption of the Rubidium atoms, its temperature dependance and Doppler broadening.

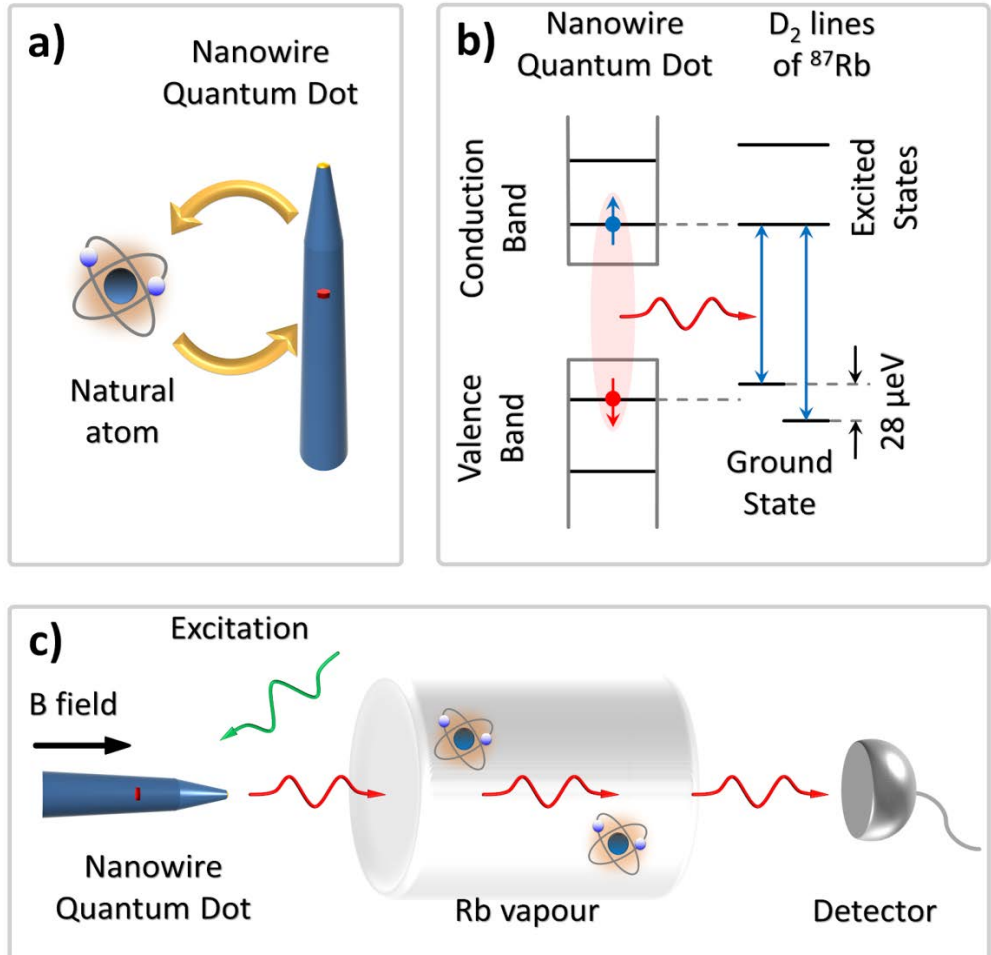

**Figure S11 | Nanowire quantum dot tuned to atomic resonances. a)** Schematics of a hybrid system of a nanowire quantum dot and natural atom (adapted from our previous work[8]), and **b)** their energy levels. An electron (blue dot-arrow) and a hole (red dot-arrow), which form an exciton in the quantum dot, recombine by emitting a photon (red wave arrow). The photon is then absorbed by the Rb vapor if its frequency matches one of the two transition lines (blue up-down arrows). **c)** Schematics of the experimental setup. The pump laser (green wave arrow) excites the quantum dot which emits a sequence of single photons (red wave arrows). These pass through a Rb vapor cell before being eventually detected. The energy of the emission is tuned by means of an external magnetic field.

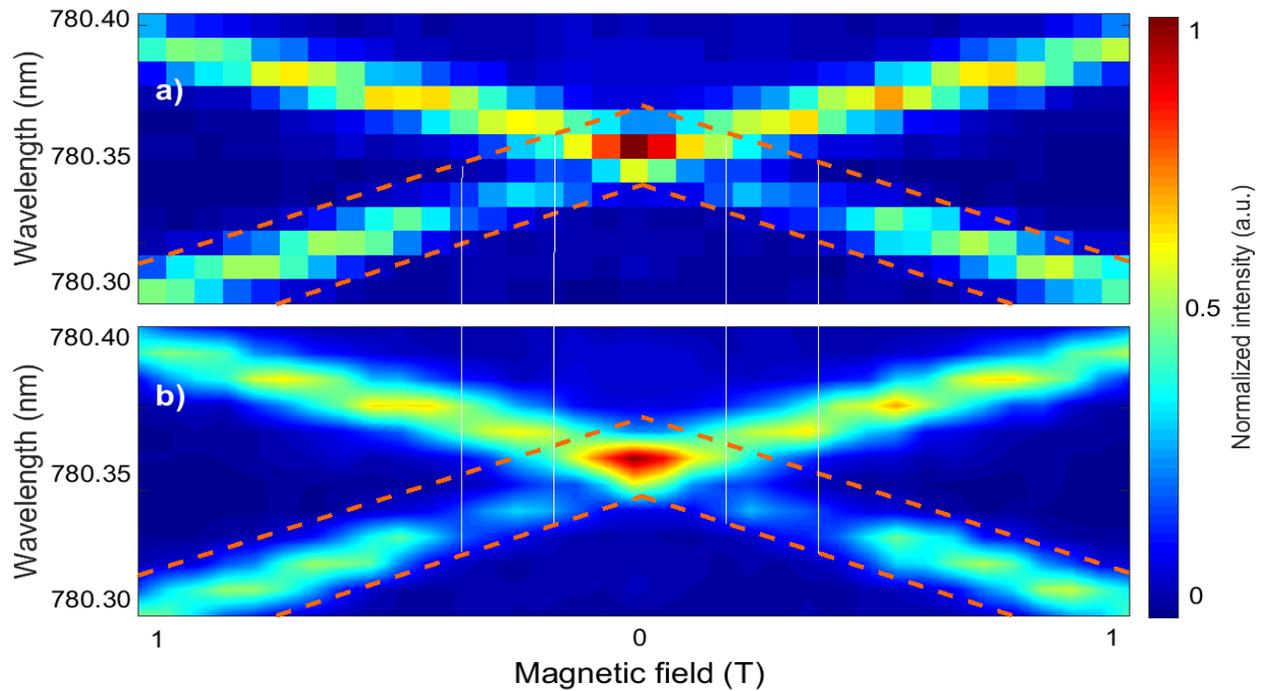

**Figure S12 | Magneto-photoluminescence. a)** Raw data of photoluminescence under applied external magnetic field for a nanowire quantum dot with emission energy close to the Rb absorption lines. **b)** 2D interpolation of the raw data, to show more clearly the absorption dips. The tracing and fitting shown in Fig. 3 of the main text use the raw photoluminescence data in a). The orange dashed lines mark the low-wavelength emission, traced for the transmission measurement in Fig. 3b of the main text. The solid white lines mark the spectral position of the $D_2$ lines of $^{87}$Rb.